\documentclass[twocolumn,showpacs,preprintnumbers,amsmath,amssymb]{revtex4}

\newcommand{\bib}{\bibitem}
\newcommand{\bea}{\begin{eqnarray}}
\newcommand{\eea}{\end{eqnarray}}
\newcommand{\beq}{\begin{equation}}
\newcommand{\eeq}{\end{equation}}
\newcommand{\non}{\nonumber}
\newcommand{\noi}{\noindent}

\newcommand{\de}{\delta}

\newcommand{\la}{\lambda}

\newcommand{\si}{\sigma}
\newcommand{\pa}{\partial}

\usepackage{graphicx,epsfig} % Include figure files
\usepackage{dcolumn} % Align table columns on decimal point
\usepackage{bm} % bold math

\begin{document}

\title{Line junction in a quantum Hall system with two filling fractions}
\author{Diptiman Sen and Amit Agarwal}
\affiliation{Center for High Energy Physics, Indian Institute of Science,
Bangalore 560 012, India}
\date{\today}

\begin{abstract}
We present a microscopic model for a line junction formed by counter or 
co-propagating single mode quantum Hall edges corresponding to different 
filling factors. The ends of the line junction can be described by two 
possible current splitting matrices which are dictated by the conditions of 
both lack of dissipation and the existence of a linear relation between the 
bosonic fields. Tunneling between the two edges of the line junction then 
leads to a microscopic understanding of a phenomenological description of 
line junctions introduced some time ago. The effect of density-density 
interactions between the two edges is considered, and renormalization group 
ideas are used to study how the tunneling parameter changes with the length 
scale. This leads to a power law variation of the conductance of the line 
junction with the temperature. Depending on the strength of the interactions 
the line junction can exhibit two quite different behaviors. Our results can 
be tested in bent quantum Hall systems fabricated recently. 
\end{abstract}

\pacs{73.43.-f, 73.43.Jn, 71.10.Pm, 73.23.-b}
\maketitle

\section{Introduction}

The recent fabrication of quantum Hall (QH) systems which have a sharp bend of
$90^0$ provides a new arena for testing theories of quantum Hall edge states
\cite{grayson1,grayson2}. By applying an appropriately tilted magnetic field,
one can create a situation in which the two faces of the
bent system are both in QH states but
they have different filling fractions $\nu_1$ and $\nu_2$; this is because the
filling fractions are governed by the components of the magnetic field 
perpendicular to the faces. If the magnetic field is sufficiently tilted, 
the two perpendicular components can even have opposite signs. Depending 
on whether $\nu_1$ and $\nu_2$ have the same sign or opposite signs, the edge 
states on the two sides of the line which separates the two QH states (called 
a line junction) propagate in opposite directions or in the same direction;
these are called counter or co-propagating edge states respectively. 

In a fractional QH system in which the two sides have the same filling 
fraction, the properties of a line junction (LJ) have been studied extensively
\cite{renn,oreg,kane1,mitra,kollar,kim,zulicke,papa,das}; they are known to 
provide a realization of a one-dimensional system of spinless interacting 
electrons with a tunable Luttinger parameter \cite{gogolin,giamarchi1,rao}.
A LJ in such a system can be formed by creating a narrow barrier which 
divides a QH liquid such that there are chiral edge states flowing on the two 
sides of the barrier \cite{kang,yang1,yang2,roddaro1,roddaro2}. In general
the edges interact with each other through a short-range (screened Coulomb)
repulsion. A LJ is therefore similar to a non-chiral quantum wire; 
however, the physical separation between the two edges of the effective 
non-chiral wire can be controlled by a gate voltage which allows for a 
greater degree of control over the strength of the interaction between 
the edges.

It is known that a LJ can be disordered, so that the tunneling amplitude 
across the LJ can be taken to be a random variable. The disorder can drive a 
localization-delocalization transition \cite{kane1}; the scaling dimension of 
the tunneling operator and therefore the occurrence of the transition 
generally depends on the strength of the density-density interaction.

Novel metallic and insulating states have been observed for a LJ 
in a bent QH system in which the filling fraction is the same on the
two sides \cite{grayson2}. The results of Ref. \cite{kane1} have
been used to understand these states. It would clearly be interesting to
extend this analysis to the case in which the two sides of the LJ have 
different filling fractions for which experimental results are expected
to be available in the near future. 

In this paper, we develop a microscopic model for a LJ between two QH states 
with different filling fractions; our model will combine ideas from several 
earlier papers. For reasons discussed below, we will work in the regime where 
the thermal decoherence length $L_T$ is much smaller than both the length $L$
of the LJ and the scattering mean free path $L_m$. We consider simple quantum
Hall states on the two sides of the LJ so that the each edge consists of only 
one chiral mode; this will happen if the filling fractions on the two sides 
$\nu_1, ~\nu_2$ are given by the inverses of odd integers like $1,3,5,\cdots$. 
In Sec. II, we discuss the idea of a current splitting matrix $S$ for a system 
with two incoming and two outgoing edges. On general grounds, such a matrix
is described by a single parameter $t$ called the scattering coefficient;
this parameter was phenomenologically introduced in Refs. \cite{wen1,halperin}.
The main aim of our work will be to provide a microscopic model for the
origin of the parameter $t$, and then to understand how $t$ varies with
the length scale or the temperature. The microscopic model will be 
developed in two stages. First, in Sec. III, we
introduce a current splitting matrix $S$ at each end of the LJ 
(described by the points $x=0$ and $x=L$). We show that the requirement
that the current splitting matrix should not lead to any dissipation exactly 
at the end leads to only two possibilities for the matrix $S$; the forms
of $S$ depend on the values of $\nu_1, ~\nu_2$. It turns out, interestingly, 
that exactly the same two possibilities for $S$ arise if we demand that the 
bosonic fields describing the chiral edge modes should be related to each 
other in a linear way. Next, in Sec. IV, we introduce the possibility 
of tunneling from a point on one edge of the LJ to the corresponding point on 
the other edge; this is described by a tunneling conductance per unit length 
$\si$ which can depend on the location of the tunneling point $x$. For $L_T 
\ll L_m$, a kinetic equation description \cite{kane2} of tunneling leads to 
a combined current splitting matrix $S_{LJ}$ for the LJ as a whole which 
depends on $\si$, $L$ and the $S$ matrices at the two ends of the LJ. We 
then turn to the temperature dependence of $S_{LJ}$ in Sec. V. We consider 
density-density interactions between the two edges of the LJ and allow for 
tunneling with a random strength between the edges. We then 
use renormalization group (RG) ideas to study how $\si$ varies with the 
temperature \cite{kane1}. In a certain regime ($L_T \ll L_m \ll L$), the 
variation turns out to be given by a power law, where the power depends on the
strength of the interaction between the two edges. Finally, the conductance 
of the LJ can be related to the matrix $S_{LJ}$. This combination of ideas 
thus gives a complete microscopic understanding of the conductance of 
the LJ, including its dependence on the temperature and length. In Sec. VI,
we discuss how our results can be experimentally tested in QH systems
with two different filling fractions. We summarize our results and 
discuss possible extensions of our work in Sec. VII.

\section{Current splitting matrix}

The main aim of our work will be to develop a model for the current splitting
matrix for a system with a line junction. To see what this matrix means, 
consider the systems shown in Fig. 1. In both the systems, the 
currents (voltages) in the two incoming edges are denoted as $I_1$ ($V_1$)
and $I_2$ ($V_2$), while the currents (voltages) in the two outgoing edges 
are denoted as $I_3$ ($V_3$) and $I_4$ ($V_4$). Here $1$ and $3$ denote the
edges of a QH system with filling fraction $\nu_1$, while $2$ and $4$ denote 
the edges of a system with filling fraction $\nu_2$. In the linear response 
regime and assuming equilibration, the currents and voltages on a QH edge 
are related as 
\beq I_{i} ~=~ \frac{e^2}{h} ~\nu_i ~V_{i}. \label{iv} \eeq

\begin{figure}[htb] \begin{center}
\includegraphics[width=1.0 \linewidth]{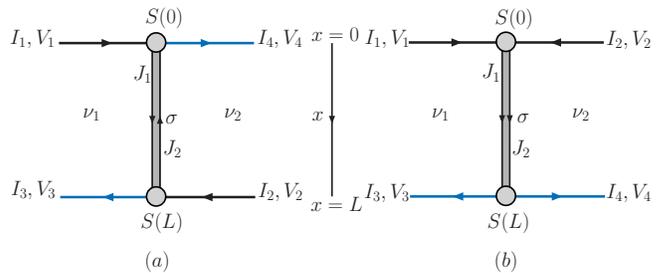} \end{center}
\caption{Schematic picture of a line junction with (a) counter propagating 
and (b) co-propagating modes.} \end{figure}

We expect that the outgoing currents should be related to the incoming ones
by a real matrix denoted as $S_{LJ}$,
\beq \left( \begin{array}{c} I_3 \\
I_4 \end{array} \right) ~=~ S_{LJ} ~\left( \begin{array}{c} I_1 \\
I_2 \end{array} \right). \label{smat} \eeq
This relation must be consistent with two general conditions:

\noi (i) current conservation, which implies that each column of $S$ should 
add up to 1, and

\noi (ii) if the incoming voltages $V_1$ and $V_2$ are equal to each other, 
the outgoing voltages $V_3$ and $V_4$ should be equal to the same quantity.

\noi Combined with Eq. (\ref{iv}), these two conditions allow a general 
current splitting matrix of the form
\beq S_{LJ} ~=~ \left( \begin{array}{cc} 1 - \frac{2 t \nu_2}{\nu_1+\nu_2} &
\frac{2 t \nu_1}{\nu_1+\nu_2} \\
\frac{2 t \nu_2}{\nu_1+\nu_2} & 1-\frac{2 t \nu_1}{\nu_1+\nu_2} \end{array}
\right), \label{generals} \eeq
where the real parameter $t$ is called the scattering coefficient
\cite{wen1,halperin}; $t=0$ represents minimum tunneling and $t=1$
maximum tunneling between the two QH fluids. 

Next, we consider the power dissipated by the system. This is given by
the difference of the incoming and outgoing power, namely,
\bea P &=& \frac{1}{2} ~[I_1 V_1 ~+~ I_2 V_2 ~-~ I_3 V_3 ~-~ I_4 V_4] \non \\
&=& \frac{e^2}{2h} ~[\nu_1 V_1^2 ~+~ \nu_2 V_2^2 ~-~ \nu_1 V_3^2 ~-~
\nu_2 V_4^2] \non \\
&=& \frac{e^2}{h} ~\frac{2\nu_1 \nu_2}{\nu_1 + \nu_2} ~t(1-t)~ (V_1 - V_2)^2.
\eea
The condition that $P \ge 0$ requires that $0 \le t \le 1$; no power is 
dissipated if $t=0$ or 1. For any value of $\nu_1, \nu_2$ and a given voltage 
difference $V_1 - V_2$, the maximum power is dissipated when $t=1/2$. 
Curiously, we note that $det ~(S_{LJ}) = 1-2t$, and vanishes at $t=1/2$.

Using Eq. (\ref{iv}), we can rewrite Eqs. (\ref{smat}-\ref{generals}) as
\bea V_3 ~-~ V_1 &=& \frac{2t\nu_2}{\nu_1+\nu_2} ~(V_2-V_1), \non \\ 
V_4 ~-~ V_2 &=& \frac{2t\nu_1}{\nu_1+\nu_2} ~(V_1-V_2). \eea
If $\nu_1 \ne \nu_2$ and $t$ lies in the range $(\nu_1+\nu_2)/(2 max (\nu_1,
\nu_2)) < t < 1$, we see that $V_3$ or $V_4$ can be higher than $max 
(V_1,V_2)$ or lower than $min (V_1,V_2)$. The system can therefore act as 
a dc step-up transformer \cite{wen1,halperin}.

\section{End of a line junction}

In this section, we will consider a current splitting matrix for each end of 
the LJ. We take each end to be a point where four edges meet, two of them 
incoming and two outgoing. One incoming and one outgoing edge is associated 
with a filling fraction $\nu_1$ and the other incoming and outgoing edge is
associated with filling fraction $\nu_2$ as shown in Fig. 2. For 
simple filling fractions $\nu_i$ given by the inverse of an odd integer,
each edge is associated with a single chiral boson as follows. Taking 
the coordinate on an edge to go from $x=0$ to $x=\infty$ ($-\infty$) for 
an outgoing (incoming) edge respectively, the Lagrangian is given by
\bea {\cal L} &=& \sum_{i=1}^2 ~[\frac{1}{4\pi \nu_i} ~\int_0^\infty dx ~\pa_x
\phi_{iO}~ (- \pa_t - v_i \pa_x) ~\phi_{iO} \non \\
& & ~~~+ \frac{1}{4\pi \nu_i} ~\int_{-\infty}^0 dx ~\pa_x \phi_{iI} ~
(- \pa_t - v_i \pa_x)~ \phi_{iI} ], \label{lag1} \eea
where $i$ labels the wire, $v_i$ denotes the velocity, and the outgoing
(incoming) fields are denoted as $\phi_{iI}$ ($\phi_{iO}$) respectively.

\begin{figure}[htb] \begin{center}
\includegraphics[width=0.5 \linewidth]{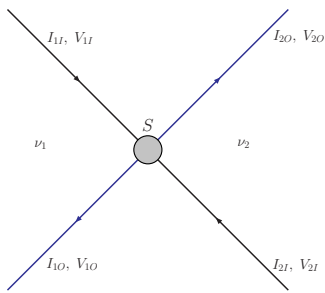} \end{center}
\caption{A meeting point of two incoming and two outgoing edges corresponding
to two different QH filling fractions.}
\end{figure}

If $\nu_i$ is the filling fraction associated with edge $i$, the 
quasi-electron and electron annihilation operators are given by $\psi_{i,qe}
\sim \eta_i e^{i \phi_i}$ and $\psi_{i,el} \sim \chi_i e^{i\phi_i /\nu_i}$
respectively, where $\eta_i$ and $\chi_i$ are the Klein factors for 
quasi-electrons and electrons respectively. The density fields canonically 
conjugate to $\phi_{iO/I}$ are given by $\rho_{iO/I} = (1/2\pi) \pa_x
\phi_{iO/I}$, so that
\bea [\rho_{iO/I} (x), \rho_{jO/I} (y)] &=& ~-~ i ~\de_{ij} 
\frac{\nu_i}{2\pi} ~\pa_x\de (x-y) \label{comm1} \eea
for points $x,y$ both lying on outgoing (incoming) edges labeled $i,j$. On the
outgoing (incoming) edge $i$, the outgoing (incoming) current is given by 
$j_{iO/I} = - (1/2\pi) \pa_t \phi_{iO/I}$. Hence current conservation implies
that $\sum_{i=1}^2 [\phi_{iO} - \phi_{iI}]_{x=0} = 0$. This implies that the
fields must be related at $x=0$ as 
\beq \left( \begin{array}{c} \phi_{1O} \\
\phi_{2O} \end{array} \right) ~=~ S ~\left( \begin{array}{c} \phi_{1I} \\
\phi_{2I} \end{array} \right), \label{outin} \eeq
where the current splitting matrix $S$ is real and each of its columns add 
up to 1.

Let us now decompose the fields at time $t=0$ as
\bea \phi_{iO} &=& \int_0^\infty ~\frac{dk}{k} ~[ b_{iOk} ~e^{ikx} ~+~ 
b_{iOk}^\dag ~e^{-ikx}], \non \\
{\rm and} ~~\phi_{iI} &=& \int_0^\infty ~\frac{dk}{k} ~[ b_{iIk} ~e^{ikx} ~
+~ b_{iIk}^\dag ~e^{-ikx}], \label{exp1} \eea
where the bosonic creation and annihilation operators must satisfy the 
commutation relations 
\beq [ b_{iOk}, b_{jOk'}^\dag ] ~=~ [ b_{iIk}, b_{jIk'}^\dag ] ~=~ \de_{ij}
\nu_i k \de (k - k') \label{comm2} \eeq
in order to satisfy Eq. (\ref{comm1}). If we now demand that the commutation 
relation in Eq. (\ref{comm2}) must be consistent with the relation in 
(\ref{outin}), we see that the matrix $S$ must satisfy
\beq S ~\left( \begin{array}{cc} \nu_1 & 0 \\
0 & \nu_2 \end{array} \right) ~S^T ~=~ \left( \begin{array}{cc} \nu_1 & 0 \\
0 & \nu_2 \end{array} \right). \label{cond1} \eeq
Using the condition of current conservation, namely, that the columns of $S$ 
should add up to 1, we find that Eq. (\ref{cond1}) implies that $S$ can only 
take two possible values, namely,
\bea S_0 &=& \left( \begin{array}{cc} 1 & 0 \\
0 & 1 \end{array} \right), \non \\
{\rm and} ~~ S_1 &=& \frac{1}{\nu_1 + \nu_2} ~\left( \begin{array}{cc} \nu_1 
- \nu_2 & 2 \nu_1 \\
2 \nu_2 & \nu_2 - \nu_1 \end{array} \right). \label{s01} \eea
Note that both these matrices satisfy $S^2 = I$. For the special case of 
$\nu_1 = \nu_2$, the second matrix reduces to
$S_1 = \left( \begin{array}{cc} 0 & 1 \\
1 & 0 \end{array} \right)$.

The power dissipated at the point $x=0$ is given by
the difference of the incoming power $(1/2) (I_{1I} V_{1I} + I_{2I} V_{2I})$
and the outgoing power $(1/2) (I_{1O} V_{1O} + I_{2O} V_{2O})$. Using Eq. 
(\ref{iv}), we find that the condition that no power is dissipated at $x=0$ 
is equivalent, in terms of the bosonic fields, to the relation $\sum_{i=1}^2
[\phi_{iI}^2 /\nu_i - \phi_{iO}^2 /\nu_i] = 0$. This implies that 
\beq S^T ~\left( \begin{array}{cc} 1/\nu_1 & 0 \\
0 & 1/\nu_2 \end{array} \right) ~S ~=~ \left( \begin{array}{cc} 1/\nu_1 & 0 \\
0 & 1/\nu_2 \end{array} \right). \label{cond2} \eeq
This is the same condition as Eq. (\ref{cond1}) since $S^2 = I$, and we 
therefore obtain the same solutions as in Eq. (\ref{s01}). We thus see that 
the conditions of zero power dissipation and a linear relation between the 
bosonic fields at the point $x=0$ are equivalent to each other; both of them 
imply that the variable $t$ appearing in the current splitting matrix 
(Eq. (\ref{generals})) at the point $x=0$ must be equal to 0 or 1.

In the next section, we will consider a LJ with either counter or 
co-propagating edges as shown in Fig. 1. We will assume that each end of the 
LJ (i.e., the points at $x=0$ and $L$) is associated with one of the matrices
$S_0$ or $S_1$ given in Eq. (\ref{s01}); there
are therefore four different possibilities for the two ends taken together.
Whether one should introduce the matrix $S_0$ or $S_1$ at each end of the LJ 
depends on the physical situation at that end. If there is a large potential 
barrier there which widely separates the QE fluids with filling fractions 
$\nu_1$ and $\nu_2$ (and therefore minimizes the possibility of tunneling), 
or equivalently, if the edges meet at the end with sharp boundaries, we 
should choose the matrix $S_0$. On the other hand, if the edges meet at a 
point with adiabatic (smooth) boundaries (with the two quantum Hall fluids 
having a greater degree of contact which allows for larger tunneling), then
the matrix $S_1$ should be chosen \cite{halperin}.

\section{Kinetic equation approach}

We now study what happens inside the LJ away from the ends. We will do this
using a simple kinetic equation approach \cite{kane2}. We assume that we are
in a steady state and there is local equilibrium at each point $x$ of the LJ.
In a steady state, the density $\rho_i (x)$ is independent of time at each 
point $x$; here $i=1,2$ denotes the edges on the two sides of the LJ.
By the equation of continuity, the currents on 
the two edges $J_1 (x)$ and $J_2 (x)$ can change with $x$ only if there is a 
current flow from one edge to the other. If there is a tunneling conductance 
per unit length given by $\si (x)$, the current flow from one edge to the 
other is given by $\si (x)$ multiplied by the potential difference between 
the two edges at the point $x$. Assuming local equilibrium, the potential at 
any point of a QH edge is related to the current as $V(x) = (h/\nu e^2)
J(x)$. We thus obtain a differential equation for the currents $J_1 (x)$ 
and $J_2 (x)$. To solve this equation, it is convenient to separately discuss 
the cases of LJs with counter and co-propagating modes shown in Figs. 1 (a) 
and (b) respectively.

\subsection{Counter propagating modes}

In the situation shown in Fig. 1 (a), we find that the currents $J_i (x)$
satisfy the equations
\beq \pa_x J_1 ~=~ \pa_x J_2 ~=~ \frac{\sigma~ h}{e^2} ~\left( 
\frac{J_2}{\nu_2} ~-~ \frac{J_1}{\nu_1} \right). \label{diff1} \eeq
Note that $J_1(x)-J_2(x)$ is constant along the LJ as one expects by current
conservation. If we assume that $\si$ is independent of $x$, we can solve
the above equations to obtain
\bea & & \left[ \begin{array}{c} J_1(x) \\ 
J_2(x) \end{array} \right] ~=~ \frac{1}{\nu_2-\nu_1} \times \non \\
& & \left[ \begin{array}{cc} \nu_2 e^{-x/l_c} - \nu_1 & \nu_1 (1 - 
e^{-x/l_c}) \\
- \nu_2 (1 - e^{-x/l_c}) & \nu_2 - \nu_1 e^{-x/l_c} \end{array} \right]
\left[ \begin{array}{c} J_1(0) \\ 
J_2(0) \end{array} \right], \label{j12x1} \eea
where $1/l_c ~=~ \frac{\sigma h}{e^2} \left(\frac{1}{\nu_1} ~-~ 
\frac{1}{\nu_2} \right)$. [If $\si$ varies with $x$, the term $x/l_c$ 
appearing in the exponentials in Eq. (\ref{j12x1}) has to be replaced by 
$\frac{h}{e^2} \left(\frac{1}{\nu_1} ~-~ \frac{1}{\nu_2} \right) ~
\int_0^x dx' \si (x')$.] 

Now $J_i (0)$ and $J_i (L)$ are related to the incoming and outgoing currents 
$I_i$ by the current splitting matrices at the ends of the LJ at $x=0$ and 
$L$. Namely,
\bea \left( \begin{array}{c} J_1 (0) \\
I_4 \end{array} \right) &=& S (0) ~\left( \begin{array}{c} I_1 \\
J_2 (0) \end{array} \right), \non \\
{\rm and} ~~\left( \begin{array}{c} I_3 \\
J_2 (L) \end{array} \right) &=& S (L) ~\left( \begin{array}{c} J_1 (L) \\
I_2 \end{array} \right). \label{slj} \eea
Using Eqs. (\ref{j12x1}-\ref{slj}), the outgoing currents can be be expressed 
in terms of the incoming currents as
\bea \left( \begin{array}{c} I_3 \\ 
I_4 \end{array} \right) &=& S_{LJ} ~\left( \begin{array}{c} I_1 \\ 
I_2 \end{array} \right), \non \\
{\rm where} ~~S_{LJ} &=& \left( \begin{array}{cc} 1 - \frac{2t \nu_2}{\nu_1
+\nu_2} & \frac{2t \nu_1}{\nu_1+\nu_2} \\ 
\frac{2t \nu_2}{\nu_1+\nu_2} & 1-\frac{2t \nu_1}{\nu_1+\nu_2} \end{array} 
\right), \label{eq:wen} \eea
and $t$ is now the scattering coefficient of the LJ as a whole.

\begin{table}[h] \begin{center} \begin{tabular}{|c|c|c|c|c|} \hline
$S(0)$ & $S(L)$ & $t$ & $t(L/l_c \to 0)$ & $t(L/l_c \to \infty)$ \\ \hline
$S_0$ & $S_0$ & $\frac{\nu_1+\nu_2}{2} \frac{1-e^{-L/l_c}}{\nu_2
-\nu_1e^{-L/l_c}}$ & $0$ & $\frac{\nu_1+\nu_2}{2 ~max (\nu_1,\nu_2)}$ \\ \hline
$S_0$ & $S_1$ & $\frac{\nu_1+\nu_2}{2} \frac{1+e^{-L/l_c}}{\nu_2+
\nu_1e^{-L/l_c}}$ & $1$ & $\frac{\nu_1+\nu_2}{2~ max (\nu_1,\nu_2)}$ \\ \hline
$S_1$ & $S_0$ & $\frac{\nu_1+\nu_2}{2} \frac{1+e^{-L/l_c}}{\nu_2+
\nu_1e^{-L/l_c}}$ & $1$ & $\frac{\nu_1+\nu_2}{2~ max (\nu_1,\nu_2)}$ \\ \hline
$S_1$ & $S_1$ & $\frac{\nu_1+\nu_2}{2} \frac{1-e^{-L/l_c}}{\nu_2-
\nu_1e^{-L/l_c}}$ & $0$ & $\frac{\nu_1+\nu_2}{2 ~max (\nu_1,\nu_2)}$ \\ \hline
\end{tabular} \end{center}
\caption{The scattering coefficient $t$ for the four possible choices of $S$ 
matrices at the ends of the LJ, for $\nu_1 \ne \nu_2$.} \end{table}

For the four different choices of $S(0)$ and $S(L)$ in terms of the two
possible current splitting matrices $S_0$ and $S_1$ in Eq. (\ref{s01}), we 
find that the scattering coefficient $t$ is given by the expressions in 
Table I. We see that depending on the choice of the matrices at the ends
of the LJ, $t$ lies in one of the two ranges $[0,(\nu_1+\nu_2)/(2 max 
(\nu_1,\nu_2))]$ and $[(\nu_1+\nu_2)/(2 max (\nu_1,\nu_2)),1]$. Note that
we need to have the non-trivial current splitting matrix $S_1$ at one of the
ends of the LJ in order to have $t$ lie in the range $[(\nu_1+\nu_2)/(2 
max (\nu_1,\nu_2)), 1]$ where the system can act as a step-up transformer.

For the special case $\nu_1 = \nu_2 = \nu$, we have to do a separate analysis 
of Eq. (\ref{diff1}) since $1/l_c = 0$. Using the same procedure as described
above, we find that the scattering coefficient is given by Table II.

\begin{table}[h] \begin{center} \begin{tabular}{|c|c|c|c|c|} \hline
$S(0)$ & $S(L)$ & $t$ & $t(L \to 0)$ & $t(L \to \infty)$ \\ \hline
$S_0$ & $S_0$ & $\frac{\sigma Lh/(\nu e^2)}{1+\sigma Lh/(\nu e^2)}$ & $0$ & 
$1$ \\ \hline
$S_0$ & $S_1$ & $1$ & $1$ & $1$ \\ \hline
$S_1$ & $S_0$ & $1$ & $1$ & $1$ \\ \hline
$S_1$ & $S_1$ & $1$ & $1$ & $1$ \\ \hline
\end{tabular} \end{center}
\caption{The scattering coefficient $t$ for the four possible choices of $S$ 
matrices at the ends of the LJ, for $\nu_1 = \nu_2 = \nu$.} \end{table}

We can relate the scattering coefficient $t$ to the two-terminal conductance 
of the LJ as 
follows. Following Fig. 1 (a), let us consider a situation in which $V_2 = 0$
and therefore $I_2 = 0$. Eq. (\ref{eq:wen}) then implies that the current
along the LJ, $I_1 - I_4 = I_3$, is related to the potential difference 
across the LJ, $V_1 - V_2$, as
\beq G_{LJ} ~\equiv ~ \frac{I_1 - I_4}{V_1 - V_2} ~=~ \frac{e^2}{h} ~\nu_1 ~
( 1 ~-~ \frac{2t\nu_1}{\nu_1 + \nu_2}). \eeq
Thus a measurement of the conductance of the LJ, $G_{LJ}$, gives the value 
of $t$. For the special case $\nu_1 = \nu_2 = \nu$, this reduces to the 
expression $G_{LJ} = (\nu e^2/h)/(1+\sigma Lh/(\nu e^2))$, where we have 
used the first line of Table II; this agrees with the expression for
the two-terminal conductance given in Ref. \cite{kane1}.

\subsection{Co-propagating modes}

We can repeat the above analysis for the case in which the two edges of the
LJ propagate in the same direction as shown in Fig. 2 (b). We now find that
the currents satisfy the equations
\beq \pa_x J_1 ~=~ - ~\pa_x J_2 ~=~ \frac{\sigma h}{e^2}~ \left(
\frac{J_2}{\nu_2} ~-~ \frac{J_1}{\nu_1} \right). \label{diff2} \eeq
Note that $J_1(x)+J_2(x)$ is constant along the edge. If we assume the
tunneling conductance $\si$ to be independent of $x$, we obtain
\bea & & \left[ \begin{array}{c} J_1(x) \\
J_2(x) \end{array} \right] = ~\frac{1}{\nu_2+\nu_1} \times \non \\
& & \left[ \begin{array}{cc} \nu_2 e^{-x/l_c} + \nu_1 & \nu_1 (1 - 
e^{-x/l_c}) \\
-\nu_2 (1 - e^{-x/l_c}) & \nu_2 + \nu_1 e^{-x/l_c} \end{array} \right]~
\left[ \begin{array}{c} J_1(0) \\ 
J_2(0) \end{array} \right], \label{j12x2} \eea
where $1/l_c ~=~ \frac{\sigma h}{e^2} \left( \frac{1}{\nu_1} ~+~ 
\frac{1}{\nu_2} \right)$.

As before, $J_i (0)$ and $J_i (L)$ are related to the incoming and outgoing 
currents $I_i$ by the current splitting matrices at $x=0$ and $L$. We then
find that the outgoing currents are again related to the incoming currents as
in Eq. (\ref{eq:wen}), where the scattering coefficient is given in Table III.
The Table remains valid for the special case $\nu_1 = \nu_2$.

\begin{table}[h] \begin{center} \begin{tabular}{|c|c|c|c|c|} \hline
$S(0)$ & $S(L)$ & $t$ & $t(L/l_c \to 0)$ & $ t(L/l_c \to \infty)$ \\ \hline
$S_0$ & $S_0$ & $\frac{1-e^{-L/l_c}}{2}$ & $0$ & $\frac{1}{2}$ \\ \hline
$S_0$ & $S_1$ & $\frac{1+e^{-L/l_c}}{2}$ & $1$ & $\frac{1}{2}$ \\ \hline
$S_1$ & $S_0$ & $\frac{1+e^{-L/l_c}}{2}$ & $1$ & $\frac{1}{2}$ \\ \hline
$S_1$ & $S_1$ & $\frac{1-e^{-L/l_c}}{2}$ & $0$ & $\frac{1}{2}$ \\ \hline
\end{tabular} \end{center}
\caption{The scattering coefficient $t$ for the four possible choices of $S$
matrices at the ends of the LJ.} \end{table}

The results given above are illustrated in Fig. 3 for the case $\nu_1 = 1$ and
$\nu_2 = 1/3$. We have shown the dependence of the scattering coefficient $t$ 
of the LJ on the dimensionless length $L \si h/e^2$ for two different choices
of the current splitting matrices at the ends of the LJ, for the cases of
counter and co-propagating edges. For the counter propagating case, $t$ 
begins at 0 (1) for $L \to 0$ and ends at $(\nu_1+\nu_2)/(2 max (\nu_1,\nu_2))
= 2/3$ for $L \to \infty$. For the co-propagating case, $t$ begins at 0 (1) 
for $L \to 0$ and ends at $1/2$ for $L \to \infty$. In this picture, we
have ignored the temperature dependence of $\si$. In the next section, we
will see how renormalization group ideas can be used to study the
temperature dependence of $\si$ and therefore of $t$.

\begin{figure}[htb] \begin{center}
\includegraphics[width=0.9 \linewidth]{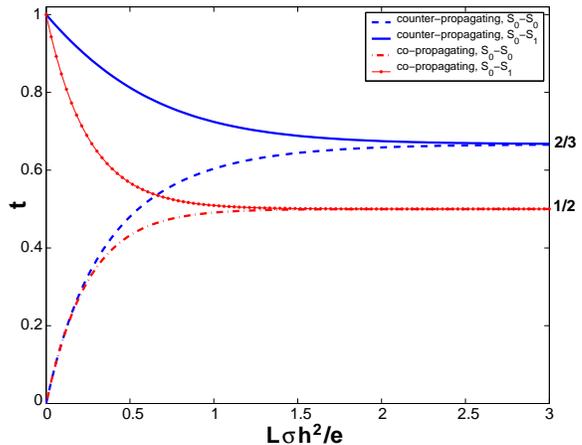} \end{center}
\caption{Scattering coefficient vs the dimensionless length of the line 
junction for two choices of the current splitting matrices at the ends $x=0$
and $L$, for the cases of counter (blue lines) and co-propagating (red lines)
edges, with $\nu_1 = 1$ and $\nu_2 = 1/3$.}
\end{figure}

\section{Disorder and Renormalization Group}

In this section, we will study the tunneling conductance $\si (x)$ in more 
detail \cite{kane1}. This arises from a tunneling amplitude $\xi (x)$ 
appearing in a Hamiltonian density
\beq {\cal H}_{tun} ~=~ \xi (x) ~\psi_1^\dag (x) \psi_2 (x) ~+~ h.c., 
\label{htun} \eeq
where $\psi_i (x)$ denotes the electron annihilation operator at point $x$ on 
edge $i$ of the LJ. The tunneling conductance $\si$ is then proportional to 
the tunneling probability $|\xi|^2$. It is believed that the presence of 
impurities near the LJ makes $\xi (x)$ a random complex variable; let us 
assume it to be a Gaussian variable with a variance $W$. The quantity $W$ 
satisfies an RG equation; to lowest order (i.e., for small $\xi$), this 
equation is given by \cite{giamarchi2}
\beq \frac{dW}{d \ln l} ~=~ (3 - 2d_t) ~W, \label{rg} \eeq
where $l$ denotes the length scale, and
$d_t$ is the scaling dimension of the tunneling operator $\psi_1^\dag \psi_2$ 
appearing in Eq. (\ref{htun}). (We will calculate $d_t$ below for both counter
and co-propagating cases). There is also an RG equation for the strength of 
the interaction between the electrons, but we can ignore that if $W$ is small.

Let us first assume that the phase decoherence length $L_T = \hbar v/(k_B T)$
(the length beyond which electrons lose phase coherence due to thermal 
smearing) is much smaller than the scattering mean free path $L_m$ of the LJ.
Successive backscattering events then become incoherent and quantum 
interference effects of disorder are absent. One can then show that $\si$ 
scales with the temperature $T$ as $T^{2d_t - 2}$ \cite{kane1}. (We note that
$\si$ is inversely proportional to the conductivity {\it along} the LJ studied
in Ref. \cite{kane1}). It therefore seems that $\si L \to 0$ as $T \to 0$ if 
$d_t > 1$. However, it turns out that this is true only if $d_t > 3/2$, i.e.,
if $W$ is an irrelevant variable according to Eq. (\ref{rg}). If $d_t > 3/2$ 
(called the metallic phase), one can simultaneously 
have $L \gg L_T$ (this is necessary to justify cutting off the RG flow at $L_T$
rather than at $L$), and $\si L \to 0$, i.e., $LT \gg 1$ and $L T^{2d_t - 2} 
\to 0$, for some range of temperatures. Within this range, one can obtain the 
scattering coefficient $t$ in Tables I - III by taking $L/l_c \sim \si L h/e^2 
\to 0$. If $d_t < 3/2$ ($W$ is a relevant variable), we have $L/L_T \sim LT 
\gg 1$ and $T^{2d_t - 3} \to \infty$; hence $\si L \sim L T^{2d_t - 2} \to 
\infty$ (we call this the insulating phase). 
We can then obtain $t$ in Tables I - III by taking $L/l_c \sim \to \infty$.
We thus see that depending on whether $d_t > 3/2$ or $< 3/2$, the parameter 
$t$ tends to quite different values as the temperature is decreased.

The above analysis breaks down if one goes to very low temperatures where $L_T
\gtrsim L$ or $L_m$. In that case, the RG flow of $W$ has to be cut off at the
length scale $L$ or $L_m$, rather than $L_T$; hence $\si$ and therefore the
scattering coefficient $t$ become independent of the temperature $T$.

In Fig. 4, we illustrate the temperature dependence of the scattering
coefficient $t$ for two choices of the current splitting matrices at the ends 
of the LJ, with $d_t = 0.8$ and $2$, for $\nu_1 = 1$ and $\nu_2 =1/3$. We have
taken the conductance $\si$ to scale as $T^{2d_t -2}$ (specifically, $L\si 
h/e^2 = T^{2d_t -2}$, where $T$ is in dimensionless units), and then 
substituted that to obtain $t$ from the first two rows of Tables I and III 
for counter and co-propagating edges respectively. For $d_t < 0.8$ (Fig. 4 
(a)), we see that $t$ approaches $2/3$ ($1/2$) as $T \to 0$ for the counter 
(co-propagating) cases respectively, for any choice of the current splitting 
matrices at the ends of the LJ. For $d_t = 2$ (Fig. 4 (b)), $t$ approaches 
$0$ ($1$) as $T \to 0$ depending on the choices of current splitting matrices 
at the ends of the LJ, regardless of whether the edges are counter or 
co-propagating. As mentioned above, these pictures become invalid when we go 
to very low temperatures where $L_T$ is not much smaller than $L$ or $L_m$.

We will now compute the scaling dimension $d_t$ of the operator $\psi_1^\dag
\psi_2$ using the technique of bosonization. It is again convenient to discuss
this for the cases of LJs with counter and co-propagating modes separately.

\begin{figure}[htb] \begin{center}
\includegraphics[width=0.9 \linewidth]{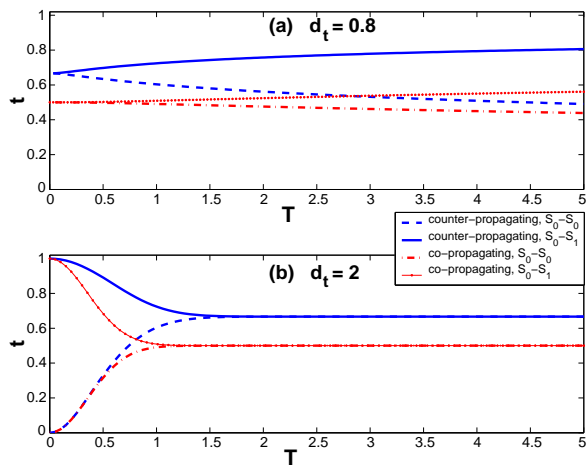} \end{center}
\caption{Scattering coefficient vs the dimensionless temperature for two 
choices of the current splitting matrices at the ends of the line junction, 
for the cases of counter (blue lines) and co-propagating edges (red lines), 
with (a) $d_t = 0.8$ and (b) $d_t = 2$, for $\nu_1=1$ and $\nu_2 =1/3$.}
\end{figure}

\subsection{Counter propagating modes} 

For the LJ shown in Fig. 2 (a), the mode on one edge goes from $x=0$ to $x=L$,
while the mode on the other edge goes in the opposite direction; let us call 
the corresponding bosonic fields $\phi_1$ and $\phi_2$ respectively. In the 
absence of density-density interactions between these modes, the Lagrangian 
is given by
\bea {\cal L} &=& \frac{1}{4\pi \nu_1} ~\int_0^L dx ~\pa_x \phi_1~ (- \pa_t -
v_1 \pa_x) ~\phi_1 \non \\
& & + ~\frac{1}{4\pi \nu_2} ~\int_0^L dx ~\pa_x \phi_2 ~(\pa_t - v_2 \pa_x) ~
\phi_2 , \label{lag2} \eea
where $v_i$ denotes the velocity of mode $i$. The bosonic fields can be 
expanded at time $t=0$ as 
\bea \phi_1 &=& \int_0^\infty ~\frac{dk}{k} ~[ b_1 ~e^{ikx} ~+~ b_1^\dag ~
e^{-ikx}], \non \\
{\rm and} ~~\phi_2 &=& \int_0^\infty ~\frac{dk}{k} ~[ b_2 ~e^{-ikx} ~+~ 
b_2^\dag ~e^{ikx}], \label{exp2} \eea
where the creation and annihilation operators satisfy the commutation relations
\beq [ b_{ik}, b_{jk'}^\dag ] ~=~ \de_{ij} \nu_i k \de (k - k'). \eeq

The electron annihilation 
operator on edge $i$ is given by $\chi_i e^{i\phi_i /\nu_i}$, $\chi_i$ being 
the electron Klein factor. The tunneling operator between the edges, 
$\psi_1^\dag \psi_2$, is therefore given by $\chi_1^\dag \chi_2 e^{i (\phi_2
/\nu_2 - \phi_1 /\nu_1)}$. (Since the edges belong to different QH systems,
quasi-particles with fractional charge cannot tunnel between the two edges
as that would change the charge of each QH system by a fractional amount).

We will again assume that $L_T \ll L_m \ll L$; therefore, two points on the LJ
which are separated by a distance much larger than $L_T$ are not related to 
each other in a phase coherent way. In particular, at all points deep inside 
the LJ, i.e., separated from the ends of the LJ at $x=0$ and $L$ by a distance
much larger than $L_T$, the bosonic fields carry no information about the 
current splitting matrices $S$ appearing at the ends. We can therefore assume 
that $\phi_1 (x)$ and $\phi_2 (x)$ are independent fields at all points $x$
except points very close to the edges. For the same reason, we can replace 
the limits of the integration in Eq. (\ref{lag2}) by $-\infty$ and $\infty$
since only fields lying within a distance of about $L_T$ from a given point 
$x$ will contribute to tunneling at that point. We can now read off the 
scaling dimension of the tunneling operator from the Lagrangian in 
(\ref{lag2}); we find that $d_t = (1/2) (1/\nu_1 + 1/\nu_2)$. For instance,
for a LJ lying between QH systems with $\nu_1, \nu_2$ equal to 1 and 1/3,
$d_t = 2$ which means that the disorder parameter $W$ is irrelevant.

We will now consider the effect of a short-range density-density interaction 
between the two edges on the scaling dimension $d_t$. The densities for
the two modes are given by $\rho_1 = (1/2\pi) \pa_x \phi_1$ and $\rho_2 = - 
(1/2\pi) \pa_x \phi_2$ respectively. Hence a repulsive interaction will
correspond to a term in the Lagrangian of the form
\bea {\cal L}_{int} &=& \frac{\la}{4\pi \sqrt{\nu_1 \nu_2}} ~\int_0^L 
dx ~\pa_x \phi_1~ \pa_x \phi_2, \label{lag3} \eea
where $\la$ is a positive number with the dimensions of velocity. The 
Hamiltonian corresponding to Eqs. (\ref{lag2}) and (\ref{lag3}) is then 
given by
\bea H &=& \int_0^\infty ~dk~ [~ \frac{v_1}{\nu_1} ~b_{1k}^\dag b_{1k} ~+~
\frac{v_2}{\nu_2} ~b_{2k}^\dag b_{2k} \non \\ 
& & ~~~~~-~ \frac{\la}{2 \sqrt{\nu_1 \nu_2}}~ (b_{2k}^\dag b_{1k}^\dag ~+~ 
b_{1k} b_{2k})~]. \eea
This can be diagonalized by a Bogoliubov transformation. We then obtain new 
bosonic fields ${\tilde \phi_1}$ and ${\tilde \phi_2}$ which have the 
velocities
\bea {\tilde v}_1 &=& \frac{1}{2} ~[\sqrt{(v_1 + v_2)^2 - \la^2} ~+~ v_1 ~-~
v_2], \non \\
{\rm and} ~~{\tilde v}_2 &=& \frac{1}{2} ~[\sqrt{(v_1 + v_2)^2 - \la^2} ~+~ 
v_2 ~-~ v_1]. \eea
The requirement of stability, ${\tilde v}_1, {\tilde v}_2 > 0$, means that we 
must have $4 v_1 v_2 > \la^2$. Finally, we can obtain the scaling dimension of
the tunneling operator $e^{i (\phi_2 /\nu_2 - \phi_1 /\nu_1)}$ after re-writing
$\phi_i$ in terms of the new fields ${\tilde \phi}_i$. We discover that
\bea d_t &=& \frac{1}{4K} \left[ (1+K^2) \left( \frac{1}{\nu_1} +
\frac{1}{\nu_2} \right) - \frac{2(1-K^2)}{\sqrt{\nu_1 \nu_2}} \right], \non \\
{\rm where} ~~K &=& \sqrt{\frac{v_1+v_2-\la}{v_1+v_2+\la}}. \label{dtk} \eea

For the special case $\nu_1 = \nu_2 = \nu$, Eq. (\ref{dtk}) gives
$d_t = K /\nu$ \cite{kane1}, while for $K=1$, we get $d_t =
1/(2\nu_1) + 1/(2\nu_2)$. It is interesting to note that for a given value 
of $\nu_1$ and $\nu_2$, $d_t$ has a non-monotonic dependence on $K$. When $K$ 
is reduced from 1 by turning on a weak repulsive interaction (i.e., $\la$ is 
small and positive), $d_t$ starts decreasing; however, $d_t$ reaches a minimum
at $K = |\sqrt{\nu_1} - \sqrt{\nu_2}|/(\sqrt{\nu_1} + \sqrt{\nu_2})$, beyond
which it starts increasing as $K$ decreases further.

\subsection{Co-propagating modes}

For the LJ shown in Fig. 2 (b), the modes on both edges go from $x=0$ to $x=L$.
Hence both modes have a Lagrangian and an expansion similar to that of the
field $\phi_1$ given in Eqs. (\ref{lag2}) and (\ref{exp2}). In the presence 
of density-density interactions, the Hamiltonian is given by
\bea H &=& \int_0^\infty ~dk~ [~ \frac{v_1}{\nu_1} ~b_{1k}^\dag b_{1k} ~+~
\frac{v_2}{\nu_2} ~b_{2k}^\dag b_{2k} \non \\
& & ~~~~~-~ \frac{\la}{2 \sqrt{\nu_1 \nu_2}}~ (b_{2k}^\dag b_{1k} ~+~ 
b_{1k}^\dag b_{2k}) ~]. \eea
This can be diagonalized by a simple rotation. The new bosonic fields have
the velocities
\bea {\tilde v}_1 &=& \frac{1}{2} ~[ v_1 ~+~ v_2 ~+~ \sqrt{(v_1 - v_2)^2 + 
\la^2}], \non \\
{\rm and} ~~{\tilde v}_2 &=& \frac{1}{2} ~[ v_1 ~+~ v_2 ~-~ \sqrt{(v_1 - 
v_2)^2 + \la^2}]. \eea
Once again stability, i.e., ${\tilde v}_2 > 0$, requires that $4 v_1 v_2 > 
\la^2$. Finally, the scaling dimension of the tunneling operator $e^{i (
\phi_2 /\nu_2 - \phi_1 /\nu_1)}$ is found to be given by 
\beq d_t ~=~ \frac{1}{2 \nu_1} ~+~ \frac{1}{2 \nu_2}, \eeq
independent of the strength of the interaction $\la$.

\section{Experimental implications}

Our results can be experimentally tested in bent QH systems such as the ones
studied in Refs. \cite{grayson1,grayson2}. A gate voltage can be used to 
control the distance between the two edges of the LJ. Making the gate voltage 
less repulsive for electrons is expected to reduce the distance between the 
edges; this should increase both the strength of the density-density 
interactions as well as the tunneling conductance \cite{kane1}. In this way, 
one may be able to vary the scaling dimension $d_t$ across the value $3/2$. 

For the case of counter propagating edges with $\nu_1 \ne \nu_2$, we have 
discussed in Sec. IV A
how $t$ can be obtained from a two-terminal conductance measurement. Our
first observation is that the value of $t$ always lies in one of two mutually 
exclusive ranges, $[0, (\nu_1+\nu_2)/(2 max (\nu_1, \nu_2))]$ or 
$[(\nu_1+\nu_2)/(2 max (\nu_1,\nu_2)),1]$; this can be seen in Table I.
Next, we saw in Sec. V that for $d_t > 3/2$, $t$ will approach either 0 or 1,
depending on which of the two ranges $t$ happens to lie in,
as the temperature is lowered (provided that $L_T << L$). On the other hand,
if $d_t < 3/2$, $t$ will approach the value $(\nu_1+\nu_2)/(2 max (\nu_1, 
\nu_2))$ from below or above, depending on which of the two ranges $t$ 
lies in, as the temperature is lowered. Finally, the rate at which the various
asymptotic values of $t$ is approached depends on the value of $d_t$; this
value is determined by $\nu_1$, $\nu_2$ and the interaction strength $\la$ 
which can be controlled by the gate voltage.

For the case of co-propagating edges with any values of $\nu_1$ and $\nu_2$, 
the value of $t$ always lies in one of two mutually exclusive 
ranges, $[0, 1/2]$ or $[1/2, 1]$; this can be seen in Table III.
For $d_t > 3/2$, $t$ will approach either 0 or 1, depending on which of the 
two ranges $t$ happens to lie in, as the temperature is lowered. If $d_t < 
3/2$, $t$ will approach the value $1/2$ from below or above as the temperature
is lowered. Unlike the case of counter propagating edges, the rate at which 
the various asymptotic values of $t$ are approached now depends on only 
$\nu_1$ and $\nu_2$, and not on the interaction strength or the gate voltage.

In the presence of interactions and disorder, $L_m$ scales with temperature 
as $T^{2-2d_t}$ and $L_T \sim T^{-1}$ \cite{kane1}. Thus throughout the 
metallic phase ($d_t>3/2$), $L_m \gg L_T$ as $T \to 0$. We note again 
that this is the regime of validity of our analysis.
 
\section{Discussion}

In this work we have developed a model for studying transport along a 
QH line junction with either counter and co-propagating modes, in the case of 
QH states for which each edge has a single chiral mode. Each end of 
the line junction is described by a current splitting matrix whose form
is severely restricted by the requirement that the bosonic fields at those
points should be linearly related to each other. We then consider the effect
of tunneling across all points of the line junction and obtain expressions
for the current splitting matrix $S_{LJ}$ of the line junction in terms of 
the filling fractions, the tunneling conductance and the length of the line 
junction. Next, the tunneling conductance is taken to be a random variable; 
its temperature dependence is obtained using renormalization group ideas.
The scaling dimension of the tunneling operator is found to depend on the
strength of the interaction between the two edges of the line junction
in the counter propagating case, but not in the co-propagating case.
Depending on the scaling dimension, the system can exhibit two different
behaviors as the temperature is decreased. For a line junction with counter 
propagating modes, one can change the behavior 
by applying a gate voltage placed above the line junction since 
such a voltage can change the effective width of the line junction and 
therefore the strength of the interactions. Our model provides a theoretical 
framework for analyzing experimental studies of the transport properties of 
line junctions in QH systems; we have discussed some experimental implications
of our results.

It would be useful to extend the analysis presented in this paper to the 
regime of very low temperature where $L_T \gtrsim L$ and $L_m$. There are 
several problems which need to be addressed in order to do this. First, the 
kinetic equation approach used in Sec. III needs to be modified in this regime
since that approach implicitly assumes that the phase decoherence length is
much smaller than the scattering mean free path of the LJ. Secondly, the 
bosonic fields on the two edges of the LJ are not independent of each other 
at low temperatures if the current splitting matrices at the ends of the LJ are
taken to be of the form $S_1$, since such a matrix mixes the bosonic fields
on all the incoming and outgoing edges if $\nu_1 \ne \nu_2$. Finally, we find
that if the phase decoherence length is larger than the length of the LJ, the 
density-density interactions by themselves lead to a non-trivial current 
splitting matrix for the system, even if the tunneling conductance $\si = 0$;
this is related to the Coulomb drag problem and will be discussed elsewhere.

Finally, we would like to mention studies of a QH system with a point-contact 
interface separating two different filling fractions \cite{sandler}, and a 
QH system with an extended constriction with the same filling fraction on 
the two sides \cite{lal}. It may be possible to extend our analysis to these 
systems as well.

\section*{Acknowledgments}
A.A. thanks A. D. Mirlin for useful discussions. A.A. thanks CSIR, India for 
financial support. We thank DST, India for financial support under the 
project SR/S2/CMP-27/2006.

\end{document}